\documentclass[11pt,a4paper]{article}
\usepackage[british]{babel}
\usepackage[utf8]{inputenc}  
\usepackage[T1]{fontenc}
\usepackage[dvips]{graphicx}
\usepackage{amssymb}
\usepackage{amsthm}
\usepackage[usenames,dvipsnames]{pstricks}
\usepackage{pst-grad}
\usepackage{pst-plot}
\usepackage{mathrsfs}
\usepackage{amsmath}
\usepackage{thmbox}
\usepackage{authblk} 
\usepackage{booktabs}
\usepackage[
  colorlinks=false,
  pdfborder={0 0 0},
  pdfauthor={Emmanuel Roubin, Jean-Baptiste Colliat},
  pdftitle={Critical probability of percolation over bounded region in N-dimensional Euclidean space},
  pdfsubject={Material science, Statistical mechanics},
  pdfkeywords={percolation theory, critical probability, excursion sets of correlated random fields},
  pdfproducer={LaTeX with hyperref package},
  pdfcreator={latex, dvips, ps2pdf},
  breaklinks=true
]{hyperref}

\def\eg {\textit{e.g.}~}

\def\x {\boldsymbol{x} }

\def\RR{ \mathbb{R} }
\def\R3{ \mathbb{R}^3 }
\def\RN{ \mathbb{R}^N }
\def\Rk{ \mathbb{R}^k }

\def\Hermite{ H\!e }
\def\Hermite{ H }

\def\LKC{ \mathcal{L} }
\def\MF{ \mathcal{M} }
\def\GMF{ \MF^{\gamma} }
\def\GMFk{ \MF^{\gamma_k} }
\def\GMFa{ \MF^{\gamma_1} }

\def\tube{ \mathcal{K} }

\def\ray{ \rho }
\def\setA{ A }

\def\ES{ E_s }

\def\lset { \kappa }

\def\uinf{ [\lset,\infty[ }
\def\tubeuinf{ [\lset-\ray,\infty[ }

\def\GRV{ X }
\def\vGRV{ \boldsymbol{\GRV} }

\def\CDF{ F }
\def\TP{ \overline{\CDF} }

\def\GRF{ g }

\def\MRF{ M }
\def\SpecMom{ \lambda_2 }
\def\normal{ \mathscr{N} }

\def\Proba{ P }
\def\Expec{ \mathbb{E} }

\def\std{ \sigma }
\def\var{ \std^2 }

\def\Lc{ l_c }
\def\Msize { l }

\def\rlinv {\beta}
\def\EC{ \chi }

\def\V{ \Phi }
\def\Vf{ \V }

\def\err{ \epsilon }
\def\errrel{ \err }

\def\pc{p_c}

\def\pl{+}
\def\mn{-}

\def\Vfc{\Vf_c}
\def\Vfcp{\Vfc^\pl}
\def\Vfcm{\Vfc^\mn}
\def\Vfci{\Vfc^\infty}
\def\lsetPp {\lset_c^\pl}
\def\lsetPm {\lset_c^\mn}

\begin{document}

%
%

\title{Critical probability of percolation over bounded region in N-dimensional Euclidean space.}

\author[1]{Emmanuel Roubin}
\author[2]{Jean-Baptiste Colliat}
\affil[1]{Laboratoire 3SR, Universit\'e Grenoble Alpes, CNRS, Grenoble INP\\ Domaine Universitaire, 38000 Grenoble Cedex, France.}
\affil[2]{Laboratoire de M\'ecanique de Lille, Universit\'e Sciences et Technologies Lille 1, CNRS, \'Ecole Centrale de Lille, Arts et M\'etiers ParisTech\\ Cit\'e Scientifique, 59655 Villeneuve d'Ascq Cedex, France.}

\maketitle

%
%

\begin{abstract}
  Following H. Tomita and C. Murakami \cite{tomita_percolation_1994} we propose an analytical model to predict critical probability of percolation. It is based on the excursion set theory which allows us to consider $N$-dimensional bounded regions. Details are given for the 3D case and statistically Representative Volume Elements are calculated. Finally generalisation to the $N$-dimensional case is made.
\end{abstract}

%
%

\section{Introduction}
\par The mass transfer properties of randomly distributed porous media are of major interest since they are involved in countless applications. Derived from chemistry \cite{flory_molecular_1941} but first expressed by \cite{broadbent_percolation_1957} as a statistical-geometry model, percolation on random stochastic patterns was used to study a fluid flow through a rock assumed to be a network of channels, randomly open or close. These problems are often treated through a lattice discretisation of space and a set of arbitrary rules (often referred as bond or site percolation) in order to define infinite cluster, which are paths for the mass transfer. Here the quantity of interest is the critical (or percolation) probability $\pc$ that defines, according to a statistical point of view, whether an infinite cluster exists or not.

\par For materials science, considering infinite size domains as well as infinite clusters is not the most convenient. In percolation theory, classical upscaling procedures use, for example, cluster statistics to overcome this limitation \cite{hunt_upscaling_1998}. Following \cite{tomita_percolation_1994} and extending it to bounded regions in space, we propose to use excursion sets of correlated Random Fields (hereafter noted RF) and their underlying predictable properties \cite{adler_new_2008} in order to tackle this problematic by the mean of topology quantification via Euler characteristic (or Euler number and hereafter noted EC).

\par The concepts of percolation and topological quantification are intrinsically different. The former raises the question of the existence of a percolating cluster of the size of the system whereas the latter is a measure of the connectivity. But, without exactly knowing why, it has been observed many times in the continuum case that the topological information carried by the EC is linked with percolation (\cite{mecke_euler_1991, mecke_morphology_1997, neuweiler_upscaling_2007}), the critical behaviour taking place when it changes sign. However, in other kind of systems such as lattices, differences can occur between critical probability predicted with percolation theory and its topological estimation \cite{neher_topological_2008, nachtrab_beyond_2015}. In addition, the present work investigates finite systems for which the phenomenon is less understood. Thus, to avoid confusion, \emph{it is knowingly assumed by the authors that, in the present context, topological measure predicts percolation}. Analysis of the link between both is out of the scope of this article but, it is the authors point of view that the strong similitudes between their results and known percolation probabilities (namely $\pc=1/2$ in 2D \cite{sykes_exact_1964} and $\pc\approx0.16$ in 3D \cite{scher_critical_1970}) justify further investigations with, for example, Monte-Carlo numerical simulations.
\par Hence a relation can be made between excursion sets and the percolation probability, using a theoretical framework. In the early nineties, it has been highlighted by \cite{tomita_percolation_1994} and \cite{okun_euler_1990}, and they deduce a critical volume fraction for infinite size problems analytically. We propose to follow the same path adding some bound corrections introduced by \cite{worsley_geometry_1996}, aptly representing finite size domains. To the authors knowledge, no analytical solution is known for these problems, leading to an expensive need of numerical resources when trying to evaluate percolation probability.

\par The outline of the paper is as follows: Section 2 presents the essential ingredients we use in terms of RFs and excursion sets. The attention is drawn on the expectation of the EC. Section 3 then turns to the evaluation of percolation probabilities over bounded regions in 3D Euclidean space. Section 4 shows how those results may lead to a new definition of the so-called Representative Volume Element (RVE), which is a major concept dealing with heterogeneous materials. Finally, Section 5 extends our result on critical probabilities for N-dimensional space.
\section{Measures of excursion sets}
\par Excursion sets are the results of a thresholding operation on a correlated RF. In the present three-dimensional case, if $\GRF:\MRF\subset\RR^3\mapsto\RR$ is such field then, for a given threshold $\lset\in\RR$, the excursion set is defined by:
\begin{equation}
  \ES=\left\{\x\in\MRF\subset\RR^3\ |\ \GRF(\x)>\lset \right\}.
  \label{eq:ES}
\end{equation}
\par
\begin{figure}[h!]
  \centering
  \noindent\includegraphics[width=\textwidth]{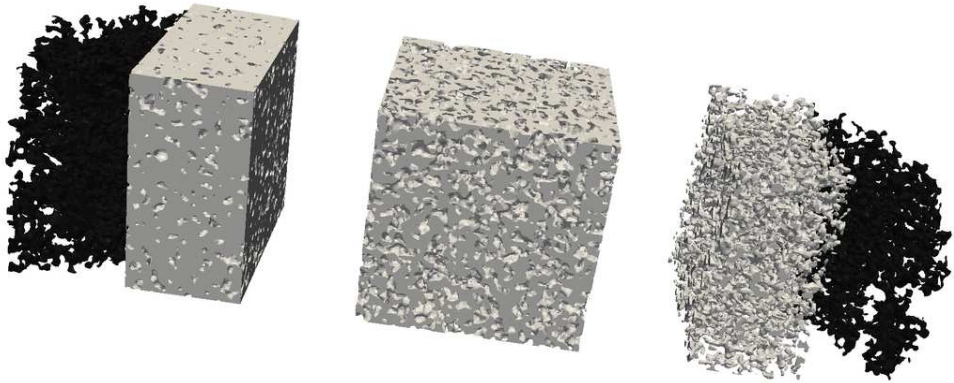}
  \caption{Excursion sets of a Gaussian RF. From left to right: low threshold and percolated cluster of empty space (black), intermediate threshold, high threshold and percolated cluster (black).}
  \label{fig:ES}
\end{figure}

 Obviously, the morphology of the excursion highly depends on the properties of the underlying RF (marginal distribution, covariance function, \dots) and the value of the threshold (see Figure~\ref{fig:ES}). However, a probabilistic link between these parameters and the morphological characteristics of the excursion is established \cite{adler_random_2007, adler_new_2008}. Among the different measures that can define a given morphology \cite{matheron_random_1975}, here attention is drawn to the volume fraction $\Vf$, which is a geometrical measure, and the EC $\EC$ which is a topological one. Obviously $\Vf$ and $\EC$ are Random Variables. Thus the link we invoke is a probabilistic one, giving the expected values ($\Expec\{\bullet\}$) of the two measures as a function of the excursion parameters.
\par Two important aspects are worth noting. First, the excursion set theory provides the model with a control of the morphologies through the analytical knowledge of the measures probability. Second, by its predictable aspect, this theory avoids the need for any actual realisation of the random morphologies and the numerical resources that it implies.
\par EC is used here to detect percolation when it changes sign (null values) \cite{okun_euler_1990}. From \cite{adler_new_2008}, its expected values are known (see equation~\eqref{eq:ELKC:0}). For a Gaussian RF $\normal(0,\var)$ of Gaussian covariance function and correlation length $\Lc$ and defined over a cube $\MRF$ of size $\Msize$ is given by:
\begin{equation}
  \Expec\{\EC\} = \left[ \frac{\rlinv^3}{\sqrt{2}\pi^2}\left(\frac{\lset^2}{\sigma^2}-1\right) + \frac{3\rlinv^2}{\sqrt{2}\pi^{3/2}}\frac{\lset}{\sigma}+\frac{3\rlinv}{\sqrt{2}\pi}\right]e^{-\lset^2/2\std^2}+\TP\left(\lset/\sigma\right), \label{eq:ELKC:0}
\end{equation}
where $\rlinv=\Msize/\Lc$ is a length ratio, $\lset$ is the threshold of equation~\eqref{eq:ES}, $\std$ the standard deviation of the normal distribution and $\TP: x \mapsto \frac{1}{\sqrt{\pi}}\int_x^{\infty}e^{-t^2}dt,$ its tail probability. Full details of the derivation of equation~\eqref{eq:ELKC:0} are given by the authors in \cite{roubin_meso-scale_2015}. It is worth noting that the asymptotic behaviour $\rlinv\rightarrow\infty$, which corresponds to infinite domains, lets equation~\eqref{eq:ELKC:0} with a single term that remains proportional to the volume times $(\lset^2-1)e^{-\lset^2}$ (for unit variance). This term, introduced by Robert Adler in 1976, is the result of differential topology where excursions could not touch the boundaries. Latter, applying the Integral geometry's Kinematic Fundamental Formula on excursions, Keith Worsley could add the three other terms (see \cite{worsley_geometry_1996} for details). These terms are the so called bound corrections which allow us to consider finite size problems.

\par Volume fraction plays directly the role of the critical volume of percolation $\Vfc$ and thus the critical probability $\pc$. Indeed, due to the definition of excursion sets of equation~\eqref{eq:ES}, it is rather straightforward that its expected value is simply the tail probability:
\begin{equation}
  \Expec\{\Vf\} = \TP\left(\lset/\sigma\right) \label{eq:ELKC:3}.
\end{equation}
\par The simple aspect of equation~\eqref{eq:ELKC:3} comes from the absence of spatial information in the measure of volume fractions. As a function of the threshold, $\Vf$ is monotonically increasing. From $1$ for infinity large negative $\lset$ to $0$ for infinitely large positive ones. However, the behaviour of the EC equation~\eqref{eq:ELKC:0} is rather more complicated. The next section explains it along with the link with percolation phenomena. More details on the excursion set theory is out of the scope of the article. As mentionned in \cite{adler_random_2007, adler_new_2008}, it has to be kept in mind that this equation still holds for:
\begin{enumerate}
\item all Gaussian related distributions,
\item all twice differentiable RFs with constant second spectral moments,
\end{enumerate}
and with it, the whole framework presented here can easily be adapted. Interested readers can consult \cite{roubin_meso-scale_2013, roubin_meso-scale_2015}.
\par The authors would like to emphasise the very similar aspect of equations~\eqref{eq:ELKC:0} and~\eqref{eq:ELKC:3} with known results on expected values of Minkowski functionals of Boolean models for infinite \cite{mecke_morphology_1997} and finite \cite{schneider_stochastic_2008} domains.

\section{Percolation criterion\label{section:percolation:3D}}
\par
\begin{figure}[h!]
  \centering
  \noindent\scalebox{1}{\input{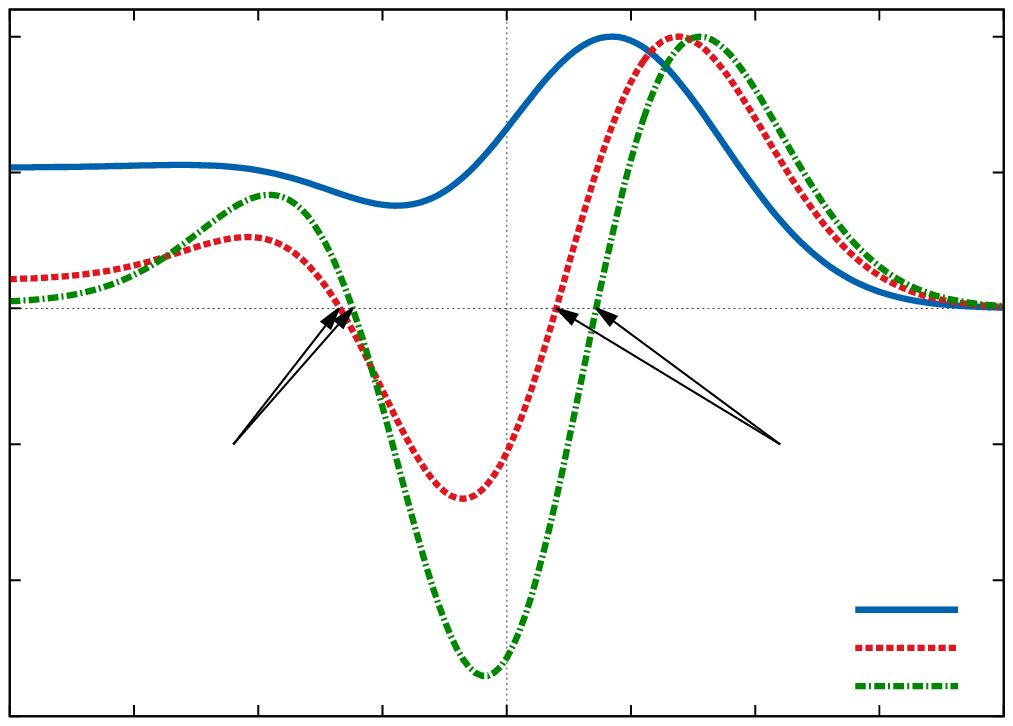}}
  \caption{Expected values of the ECs as a function of the threshold value $\lset$ for different length ratios $\rlinv$ for Gaussian RF with Gaussian covariance function. To fit in a single graph, $\EC$ is normalised to have a unitary maximal value.}
  \label{graph:EC}
\end{figure}

 The link between percolation and the EC is considered when, for variations of the threshold, the EC switches signs \cite{mecke_morphology_1997}. As depicted in Figure~\ref{graph:EC} where the expected value of the EC of equation~\eqref{eq:ELKC:0} is plotted as a function of the threshold value $\lset$ for three different length ratios $\rlinv$, most of the time it occurs two times and represents two different percolation states.
\par The two roots of $\Expec\{\EC(\lset)\}$ equation~\eqref{eq:ELKC:0} noted $\lsetPp$ and $\lsetPm$ correspond to the following situations:
\begin{description}
  \item[Percolation of the excursion at $\Expec\{\EC(\lsetPp)\}=0$]\hfill\\ By considering a high threshold value, the corresponding excursion set is made of small disconnected components. As the threshold value decreases, these components grow and others appear making the EC increase up to a certain maximum point. It corresponds to the coalescence of the biggest components. The critical point of interest $\lsetPp$ is for a threshold value just below the latter states, where the $\EC$ is null. It is stated that at this stage, the excursion set is percolated.
  \item[Percolation of the voids $\Expec\{\EC(\lsetPm)\}=0$]\hfill\\ The same reasoning can be made in order to define the other percolation point. This time, increasing thresholds starting at a very low value has to be considered. The percolation point $\lsetPm$ is also for the first null value of the EC and corresponds to the percolation of the voids.
\end{description}
\par It can be noted in Figure~\ref{graph:EC} that, for some values of $\rlinv$, percolation cannot be defined ($\rlinv=2$ for example). Since the EC is always positive, the volumes considered are too small to bring any relevant statistical information in terms of percolation states. This observation can naturally be understood by considering the asymptotic case $\rlinv\rightarrow 0$ where equation~\eqref{eq:ELKC:0} simply becomes the tail probability $\TP$ of the underlying distribution. In this specific case, correlation is infinite compared to domain size, meaning that the RF is constant leading to excursions without any other topological meaning than being completely full $\EC=1$ or completely empty $\EC=0$. This explains why the fraction volume and the EC are equivalent. Increasing $\rlinv$ brings back the spatial structure of correlated RF (and with it the differentiation between the fraction volume and EC) until the volume is representative enough to have critical behaviour, \eg null values of $\EC$. Attention is now drawn to the corresponding critical volume for percolation:
\begin{equation}
  \Vfcp=\Expec\left\{\Vf(\lsetPp)\right\} \ \ \text{and}\ \ \Vfcm=\Expec\left\{\Vf(\lsetPm)\right\}.
  \label{eq:critical-volumes}
\end{equation}

\begin{figure}[h!]
  \centering
  \noindent\scalebox{1}{\input{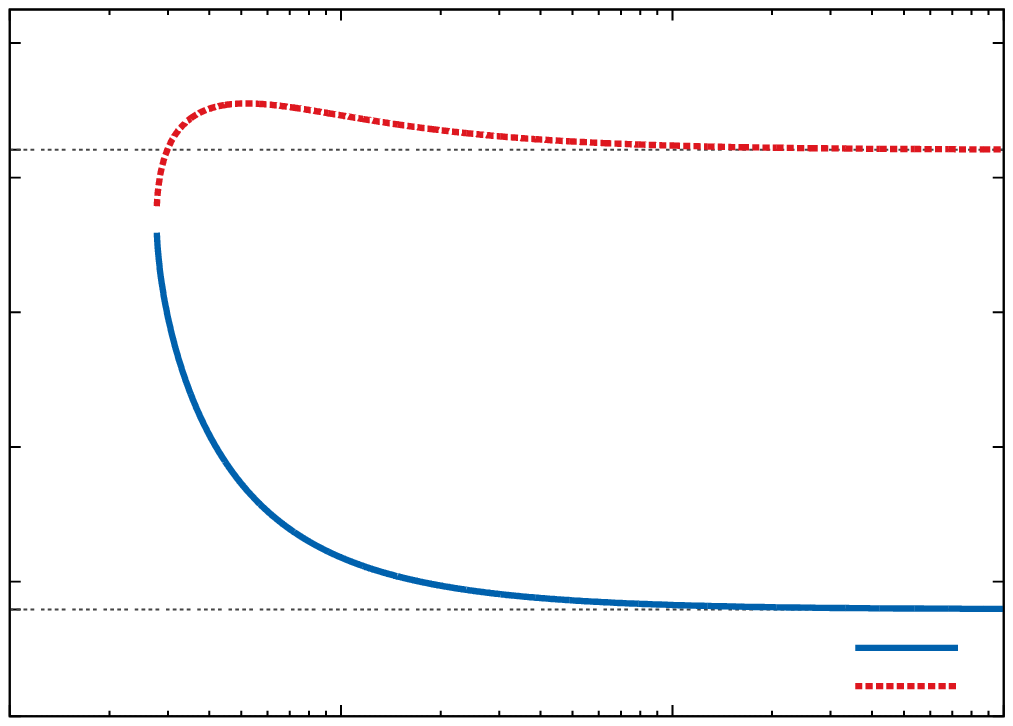}}
  \caption{Critical volumes of percolation for both excursion set ($\Vfcp$) and voids ($\Vfcm$) of Gaussian RF with Gaussian covariance function as a function of the scale ratio $\rlinv$ in a three dimensional cube.\label{graph:percolation}}
\end{figure}

Even if not taken into account in the notation, the critical volumes of equation~\eqref{eq:critical-volumes} have to be considered as expected values. Figure~\ref{graph:percolation} shows them both as a function of the length ratio $\rlinv$.
\par It can be seen that a frontier is created between two states, one with positive values of $\EC$ (outside) and the other with negative values (inside). As the topology of an excursion can be determined whether it is inside or outside, this graph can be seen as a finite size approximation of a phase diagram. The fact that the two curves go away from their asymptotic values and finally meet is a direct representation of side effects analytically taken into account by the bound corrections present in equation~\eqref{eq:ELKC:0} \cite{worsley_geometry_1996}. It is this feature that grants to the model the possibility to predict percolation over bounded regions.

\par The asymptotic values of Figure~\ref{graph:percolation} can easily be calculated. Indeed, when $\rlinv\rightarrow\infty$, the two roots of equation~\eqref{eq:ELKC:3} are $\lsetPp=\std$ and $\lsetPm=-\std$. Injecting them in equation~\eqref{eq:ELKC:3} the two critical volumes:
\begin{equation}
    \lim_{\rlinv\rightarrow \infty} \Vfcp = \TP(1) = 0.1584\dots
\end{equation}
and
\begin{equation}
  \lim_{\rlinv\rightarrow \infty} \Vfcm = \TP(-1) = 0.8413\dots
\end{equation}
\par Furthermore, from the properties of the tail probability function in this case, the symmetry between void and excursion percolation is proven by the fact that:
\begin{equation}
    \lim_{\rlinv\rightarrow \infty} \Vfcp = 1 - \lim_{\rlinv\rightarrow \infty} \Vfcm.
\end{equation}
\par These results are consistent with some values from the literature obtained first by computer simulations of the potential model ($0.17\pm 0.01$ in \cite{skal_percolation_1973}) or by speculation on lattice problems ($0.16\pm 0.02$ in \cite{zallen_physics_1983}). They are also consistent with the symmetrical assumption $\Vfcp+\Vfcm=1$. However, based on numerical simulations \cite{garboczi_geometrical_1995,rintoul_precise_1997} or analytical approximations \cite{yi_analytical_2004}, three dimensional continuum models are often based on overlapping ideal geometrical objects (sphere, ellipsoids, squares, \dots) leading to a significant variation of the results (see \cite{thorpe_sixteen-percent_2001} for a more in-depth analysis).
\par Now, still in Figure~\ref{graph:percolation}, as $\rlinv$ decreases, the critical volume fraction of the excursion set $\Vfcp$ increases. It agrees with remarks made in \cite{rintoul_precise_1997} stating that simulations in larger domains lead to smaller results. This rather unintuitive feature can be interpreted by the fact that, in a finite size domain, there are less possible paths to link two distant points than in an infinite one. Hence, a higher volume fraction is needed. Table~\ref{tab:criticalvsscaleration} shows some approximations of $\Vfcp$ and $\Vfcm$.
\begin{table}[h!]
  \centering
  \caption{Critical volumes of percolation of excursion sets of Gaussian RF with Gaussian covariance function for specific scale ratios.\label{tab:criticalvsscaleration}}  
  \begin{tabular}{c|cccccc}
    \toprule
    $\rlinv$ & $2.7766$ & $5$ & $10$ & $100$ & $1000$ & $\infty$\\
    \midrule
    $\Vfcp$ & 0.7257 & 0.3454 & 0.2367 & 0.1652 & 0.1593 & 0.1587\\
    $\Vfcm$ & 0.7509 & 0.9101 & 0.8257 & 0.8477 & 0.8420 & 0.8413\\
    $\Vfcp+\Vfcm$ & 1.4766 & 1.2555 & 1.0624 & 1.0129 & 1.0013 & 1\\
    \bottomrule
  \end{tabular}
\end{table}
\par First, it can be observed that the present prediction model possesses a limitation for small scales ($\rlinv<\rlinv_\text{lim}$ with $\rlinv_\text{lim}\approx2.7766$ see table~\ref{tab:criticalvsscaleration}) where no percolation can be defined. As mentioned in the previous section, it corresponds to an absence of root in the EC function. Then, table~\ref{tab:criticalvsscaleration} shows that for finite size domains we observe the loose of the symmetrical relationship $\Vfcp+\Vfcm\ge1$, meaning that the symmetry hypothesis (that exists on infinite size problems) does not apply anymore. Finally, when looking at the variation of the critical volumes as a function of the scales, it can be seen that the side effects play a significant role for rather small scales. Actually, as a first part of next section, a quantification of the contribution of side effects leads to the definition of morphological RVE for percolation. Then an extension of the predictive model is made for $N$-dimensional spaces.

\section{Possible extensions}
\subsection{Percolation Representative Volume Elements\label{section:RVE}} 
Dealing with heterogeneous media, traditional approaches to define RVE are based upon a set of realisations and an averaging procedure. Theoretically, if the domain $\MRF$ is a RVE for a given property, the discrepancy of the results must vanish to zero. However, these RVE are usually too large to be handled numerically. Smaller realisations are thus made and from average values over all the realisations, lower and upper bounds can be defined \cite{huet_application_1990,hazanov_order_1994}. Herein, this statistical approach is embedded in the theoretical framework that gives the excursion set expected characteristics. Hence, no actual realisation is computed.
\par Due to its monotonic shape, the RVE is defined in regards to the critical volumes of percolation of excursion set (now noted $\Vfc$ instead of $\Vfcp$), the ``reference'' value being the asymptotic value on infinite domains:
\begin{equation}
  \Vfci=\lim_{\rlinv\rightarrow \infty} \Vfc.\label{eq:perco:infinity}
\end{equation}
The relative error
\begin{equation}
	\errrel= \frac{\Vfc-\Vfci}{\Vfci}\label{eq:perco:ver-error}
\end{equation}
gives the RVE precision. Hence, a value of $\rlinv$ can be associated to a given error. Table~\ref{tab:perco:ver-error} shows that, for a given heterogeneity size $\Lc$, a region of size larger than $80$ times $\Lc$ is needed for $\errrel=0.05$ and $400$ times for $\errrel=0.01$.
\begin{table}[h!]
  \centering
  \caption{Length ratio and corresponding critical volume of $\errrel=0.01$ and $\errrel=0.05$ error RVE for percolation of excursion sets of Gaussian RF with Gaussian covariance function.\label{tab:perco:ver-error}}
  \begin{tabular}{c|ccc}
    \toprule
    $\errrel$ & $0$ & $0.01$ & $0.05$ \\
    \midrule
    $\Vfc$ & $15.87$ & $16.03$ & $16.67$ \\
    $\rlinv$ & $\infty$ & $400$ & $83.33$\\
    \bottomrule
  \end{tabular}
\end{table}
\par As far as the authors can tell, no RVE for percolation can be found in the literature. However, it can be compared to classical mechanical problems linked with percolation issues such as diffusivity phenomena or permeation in cement paste. On these matters, results are rather consistent and tend to define RVE corresponding to a scale ratio of $\rlinv=100$. Among the vast literature on that subject, in \cite{zhang_microstructure-based_2011} the authors define a cement paste RVE for water diffusivity of $100\time100\time100\,\mu m^3$ with heterogeneities represented by polydisperse spheres from $1\,\mu m$ to $50\,\mu m$. The RVE is then smaller than what we predict. Several considerations can explain this difference. First, the property of interest differs. When the latter RVE is based on mechanical property (diffusivity), the theoretical approach suggested in this article only takes into consideration the topological aspect of the morphology. Secondly, a size distribution of spheres is compared to an excursion set with one characteristic length only.

\subsection{$N$-dimensional percolation\label{section:percolation:ND}}
As presented in the previous sections, equations~\eqref{eq:ELKC:0} and \eqref{eq:ELKC:3} link the marginal distribution of the RF and its spatial structure to the geometrical and topological characteristics of the excursion. Such theoretical results come from \cite{taylor_euler_2003,taylor_gaussian_2006} in a rather more general aspect than presented in the paper. It is especially valid for RF (and thus, excursions) defined over $N$-dimensional spaces.
\par In such space, any morphology can be characterised by $N+1$ independant mathematical measures (\cite{hadwiger_vorlesungen_1957, chen_simplified_2004}). In the present case, the base made of the $N+1$ Lipschitz-Killing curvatures $\LKC_j$ is used \cite{adler_new_2008}. As a measure of Euclidean spaces and ignoring a factor of scale, they are equivalent to the more commonly used Minkowski functionals and can be seen as $j$-dimensional sizes. For example, for an $N$ dimensional hypercube $C_N=\Pi_{i=1}^N[0\ a]$, they read $\LKC_j(C_N)=\left(\begin{array}{c}N\\j\end{array}\right)a^j$, for $j=[0..N]$. In $3$D they will represent the volume, half of the surface, twice the calliper diameter and the EC, for $j=$ 3, 2, 1 and 0, respectively.
  \par As before, the necessary conditions for the RF is to be twice differentiable and have a constant second moment. Now, let $\GRF:\MRF\subset\RR^N\mapsto\RR$ be a Gaussian RF and $\SpecMom$ be its second spectral moment. If $\ES=\left\{\x\in\MRF\subset\RN\ |\ \GRF(\x)>\lset \right\}$ is the corresponding excursion set, expectation of the Lipschitz-Killing curvatures can be computed, for $j=[0..N]$, as follows \cite{adler_new_2008}:
\begin{equation}
  \Expec\{\LKC_j(\ES)\} = \sum_{i=0}^{N-j}\left(\!\!\begin{array}{c} i+j \\ i \end{array}\!\!\right)\frac{\omega_{i+j}}{\omega_{i}\omega_j}\left(\frac{\SpecMom}{2\pi}\right)^{i/2}\!\!\!\!\LKC_{i+j}(\MRF)\,\GMF_i(\lset),
  \label{eq:ELKC:general}
\end{equation}
where $\omega_j=\pi^{N/2}/\Gamma(1+N/2)$ are the $j^\text{th}$ volume of the unit ball and, carrying the information of the RF distribution, $\GMF_i$ are the $i^\text{th}$ Gaussian Minkowski functionals\footnote{Minkowski-like functionals are used here to measure the probabilistic space whereas usually Minkowski functionnals are used to measure Euclidean spaces. Herein the Lipschitz-killing curvatures do this task.}. It is worth noting that only the latter fonctionnals depend on the threshold $\lset$ (see appendix \ref{app:minkowski} for their analytical definitions and \cite{roubin_meso-scale_2013, roubin_meso-scale_2015} for more general cases). The EC is the ``$0^\text{th}$'' topological measure $\EC=\LKC_0(\ES)$, whereas the equivalent of the fraction volume is the specific $N^\text{th}$ measure $\Vf=\LKC_N(\ES)/\LKC_N(\MRF)$.
\par
\begin{figure}[h!]
  \centering
  \noindent\scalebox{1}{\input{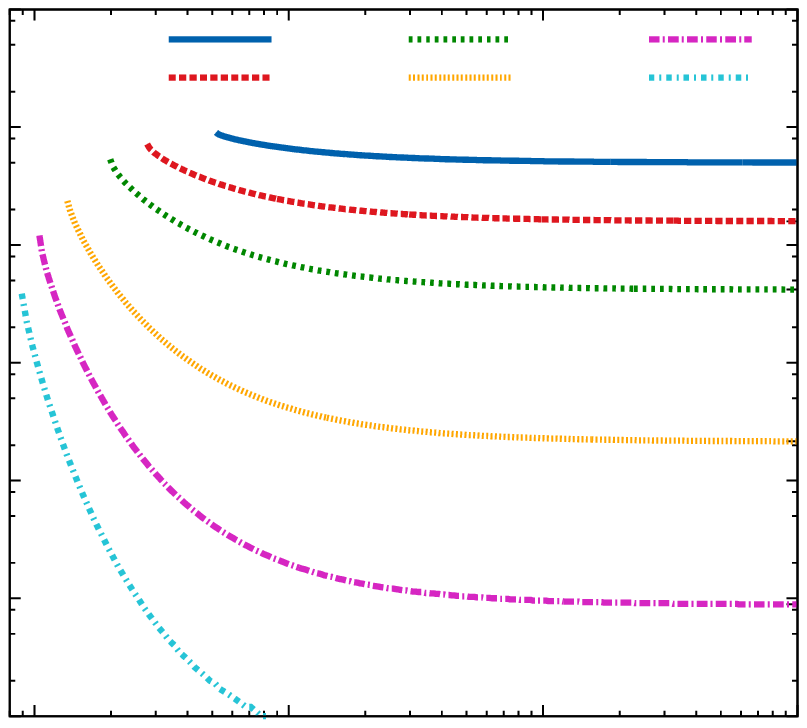}}
  \caption{Critical volume for percolation $\Vfc$ as a function of the scale ratio $\rlinv$ for hypercubes in different spatial dimensions in the case of excursion sets of Gaussian RF with Gaussian covariance function.\label{graph:percolation:lastRoot:ND}}
\end{figure}

 With this generalisation of equations~\eqref{eq:ELKC:0} and \eqref{eq:ELKC:3}, the same methodology as in section~\ref{section:percolation:3D} can be applied for any $N$. It is in this spirit that Figure~\ref{graph:percolation:lastRoot:ND} draws the critical volumes for percolation (for $N=2,3,4,6,8$ and $10$). The critical volumes are now defined for percolation of the excursion. Thus $\Vfc=\Expec\{\Vf(\lsetPp)\}$, with $\lsetPp$ being the larger root of the expected EC. 
\par Accordingly to the literature for infinitely large domains and extended to all scale ratio, the critical volumes decrease along with increasing spatial dimensions. Furthermore, increasing spatial dimensions lead to larger RVE (see section~\ref{section:RVE}). In other terms, the asymptotic behaviour is obtained for larger domains. However, it has to be kept in mind that Figure~\ref{graph:percolation:lastRoot:ND} draws only the percolation probability corresponding to the largest root $\lsetPp$. As noticed in \cite{tomita_percolation_1994}, in $N$ dimensional infinite spaces the EC has $N-1$ roots corresponding to $N$ different ``states'' of percolation and $N-1$ transition times (the EC roots). In two dimensions there is one transition time between two states, namely when it changes, for increasing thresholds, from a percolated excursion and non percolated voids to a non percolated excursion and percolated voids. In three dimensions two transitions occur: first from a percolated excursion and non percolated voids (closed porosity type of morphology) to percolated excursion and voids (open porosity type of morphology) and second from the latter state to a non percolated excursion to percolated voids (connected components type of morphology). Since it goes beyond the standard definitions it is not clear to us what the other states mean in higher dimensions. We think that, as three dimensional topological entities (handles) allows us to yield additional percolation states compare to the two dimensional case, higher topological entities can produce the same effect. However, it is because we consider only the higher EC root that we assume it detects the percolation of the excursion.

\par
\begin{figure}[h!]
  \centering
  \noindent\scalebox{1}{\input{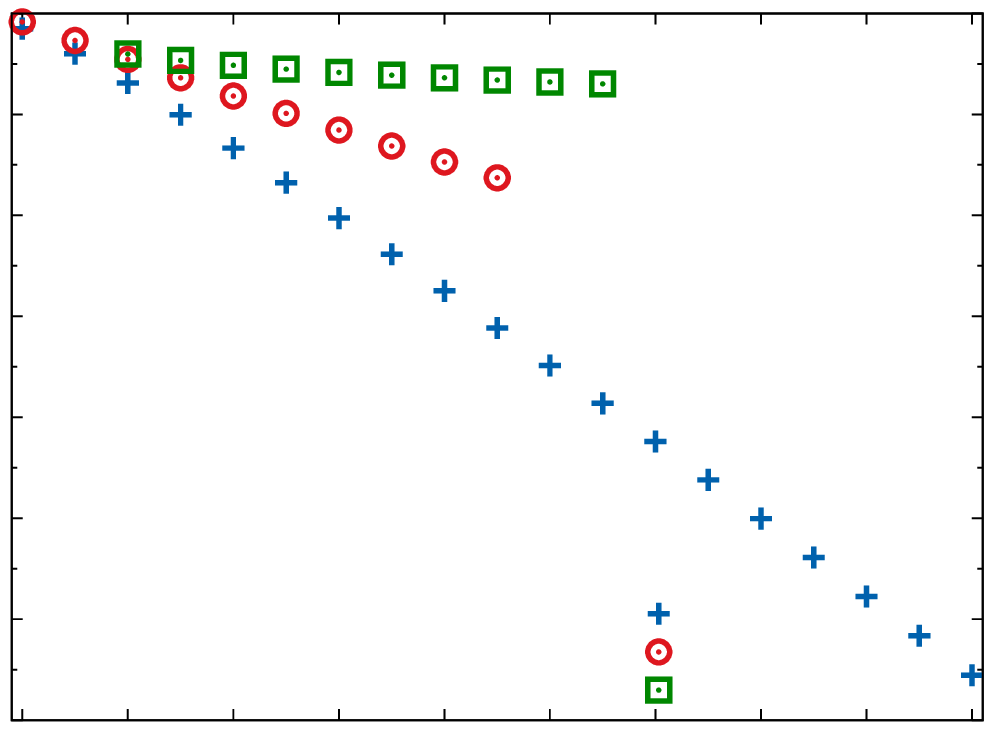}}
  \caption{Asymptotic critical volume for percolation $\Vfci$ of infinitely large excursion set ($\rlinv\rightarrow\infty$) as a function of the spatial dimension of the hyper-cubical domain definition $N$. Equivalent values are also drawn from \cite{torquato_effect_2012} and \cite{grassberger_critical_2003} for overlapping spheres and lattice bond percolation, respectively.\label{graph:percolation:inf:ND}}
\end{figure}

 Finally attention is drawn to the asymptotic behaviour of the critical volumes for very large domains. For that matter, Figure~\ref{graph:percolation:inf:ND} shows $\Vfci$ as defined in equation~\eqref{eq:perco:infinity} as a function of $N$.
\par As it was already noticeable in Figure~\ref{graph:percolation:lastRoot:ND}, it can be seen in Figure~\ref{graph:percolation:inf:ND} that, for $N=2$, the critical percolation probability of bond lattices $\pc=1/2$ \cite{kesten_percolation_1982} is retrieved. However, going further in dimensions increase drastically the difference between the models. Comparison is made here, first with results from \cite{torquato_effect_2012} where percolation threshold $\eta_c$ represents the total specific volume of overlapping hyperspheres. For the sake of comparison, the data drawn are the equivalent critical volume $\Vfc=1-e^{-\eta_c}$. Then comparison is made with bond percolation probability $\pc$ on lattice from \cite{grassberger_critical_2003}. The difference between these results comes from the rather different physical natures of the models and even the definition of percolation. However, the tendency of smaller critical percolation probability with higher dimension seems to be confirmed.

\section{Concluding remarks}
\par We propose an analytical model that determines the critical probability of percolation over bounded regions in $N$-dimensional Euclidean space through topological estimation. It is based on the probabilistic knowledge of measures of excursion sets based on correlated RF. First, the EC is used to detect percolation through its null values and then, the corresponding $N$-dimensional volume fraction is the so called critical volume $\Vfc$ that links the continuum aspect of excursions and the well known critical probability $\pc$.
\par Analysis have mainly been made through 3D examples. The model is able to retrieve results of the literature such as the critical volume $\Vfc\approx0.16$ for infinite regions. But, more importantly, since the model can predict percolation for bounded regions the percolation probability is considered as a function of a scale ratio (or domain size) and the results stated just above are their asymptotic values. This feature is used in a concrete example where statistically Representative Volume Elements for percolation have been analytically determined. An extension to $N$-dimensional spaces is proposed, showing the general aspect of this analytical framework. However, in opposition to other frameworks, this model does not provide information on the subcritical and supercritical behaviour.
\par Several lines of investigation can come out from these results. For example, an analysis between topological prediction and the critical behaviours in the case of excursion sets can be made with Monte-Carlo simulations in order to see how well does the EC predict continuum percolation. Another fruitful way is to compare the present approach with average concentrations of spanning clusters from finite-size scaling frameworks (see chapter 4.1 of \cite{stauffer_introduction_1994}) where impact of the size of the observation window can be analysed.

\appendix
\section{Gaussian Minkowski functionnals\label{app:minkowski}}
By explaining how the Minkowski functionnals are linked with the probability of the Random Field to be greater than the threshold $\GRF(\x)>\lset$, or in other words, the probability of the Random Field to be in the so-called hitting set $\GRF(\x)\in\uinf$, this appendix gives the analytical definition of the Gaussian Minkowski functionnals $\GMF_i$ in equation \eqref{eq:ELKC:general}.

\subsection{Taylor expansion of the tube Gaussian volumes}
In contrast to Lipschitz-killing curvatures, the Minkowski functionals are not intrinsic. Therefore, they depend on the measure asociated to it. Herein, the measure of a Gaussian distribution is of concern. Let $\gamma_{k}$ be such a measure in the Euclidean space $\Rk$. If $\vGRV=\{\GRV_i\}$ is a standard Gaussian vector of size $k$ in which $\GRV_i \sim \normal(0,\std^2),\ i=[1..k]$ are independent and $\setA\subset\Rk$:
\begin{equation}
  \gamma_k(\setA) = \Proba\left\{\vGRV\in \setA\right\} = \frac{1}{\std^k(2\pi)^{k/2}} \int_\setA e^{-\|\x\|^2/2\std^2} d\x.
  \label{eq:gaussian-volume}
\end{equation}
Associated to this measure, the functionals are now called Gaussian Minkowski functionals (hereafter GMFs). In a $k$-dimensional space, $k+1$ GMFs are defined, noted $\GMFk_j$, $j=[0..N]$. The main results of \cite{taylor_gaussian_2006} is to yield the following Taylor expansion of tube probability:
\begin{equation}
  \gamma_k(\tube(\setA,\ray)) = \sum_{j=0}^\infty \frac{\ray^j}{j!}\GMFk_j(\setA).
  \label{eq:taylor-expansion-gaussian-volume}
\end{equation}
for small enough $\ray$, where $\tube(\setA,\ray)$ is the tube of $\setA$ of ray $\ray$. It can be seen as an extension of the Steiner-Weyls formula, in which $\gamma_k$ represents the volume in the sense of Gaussian measure (Gaussian volume). By taking $\ray=0$, it can be concluded that the first Gaussian Minkowski functional is the Gaussian volume of $\setA$ itself, $\GMFk_0(\setA)=\gamma_k(\setA)$. Other GMFs can be identified in specific cases.

\subsection{Application to scalar-valued Gaussian Random Fields}
Even though the presented formulae work for more general cases, it suffices in this case to take $k=1$, corresponding to scalar-valued RF $\GRF:\R3\rightarrow \RR$ and excursion sets defined as:
$$\ES=\left\{\x\in\MRF\subset\RN\ |\ \GRF(\x)>\lset \right\}=\left\{\x\in\MRF\subset\RN\ |\ \GRF(\x)\in\uinf \right\}$$
It leads to several simplifications. First, the Gaussian volume of the hitting set $\uinf$ is the complementary cumulative density function (or tail distribution, noted $\TP$) of the underlying standard distribution:
\begin{equation}
  \gamma_1(\uinf) = \Proba\left\{\GRV\ge \lset\right\} = \frac{1}{\std \sqrt{2\pi}} \int_\lset^\infty e^{-x^2/\std^2} dx = \TP(\lset),
\end{equation}
furthermore, the tube of $\uinf$ can easily be defined and its tail probability linked to the tail distribution as follows:
\begin{equation}
  \tube(\uinf,\ray) = \tubeuinf \ \ \text{and} \ \ \gamma_1(\tubeuinf) = \TP(\lset-\ray).
\end{equation}
Each GMFs of \eqref{eq:taylor-expansion-gaussian-volume} can be identified by the unique standard Taylor expansion of $\TP(\lset-\ray)$ for small $\ray$, leading to:
\begin{subequations}
  \begin{align}
    \text{for } j=0  , &\ \ \GMFa_0(\uinf) = \TP(\lset), \\
    \text{for } j\ge1, &\ \ \GMFa_j(\uinf) = (-1)^j\,\frac{\text{d}^j\,\TP(\lset)}{\text{d}\lset^j} = \frac{e^{-\lset^2/2\std^2}}{\std^j\sqrt{2\pi}}\Hermite_{j-1}\left(\lset/\std\right),
  \end{align}
  \label{eq:GMFs-Gaussian-distribution}
\end{subequations}
where $\Hermite_j, j\ge0$ are the $j^\text{th}$ probabilist Hermite polynomials:
$$\Hermite_n(x) = (-1)^n\,e^{x^2/2} \frac{\text{d}^n}{\text{d}x^n}\,e^{-x^2/2}.$$

\par More general results can be find in \cite{roubin_meso-scale_2013}.

\bibliographystyle{plain}
\bibliography{biblio}

\begin{thebibliography}{10}

\bibitem{adler_random_2007}
Robert~J. Adler and Jonathan~E. Taylor.
\newblock {\em Random {Fields} and {Geometry}}, volume XVII of {\em Springer
  {Monographs} in {Mathematics}}.
\newblock Springer, New York, 2007.

\bibitem{adler_new_2008}
RobertJ. Adler.
\newblock Some new random field tools for spatial analysis.
\newblock {\em SERRA}, 22(6):809--822, 2008.

\bibitem{broadbent_percolation_1957}
S.~R. Broadbent and J.~M. Hammersley.
\newblock Percolation process, {I} and {II}.
\newblock {\em Proc. Cambridge Philos. Soc.}, 53:629 -- 645, 1957.

\bibitem{chen_simplified_2004}
Beifang Chen.
\newblock A {Simplified} {Elementary} {Proof} of {Hadwiger}'s {Volume}
  {Theorem}.
\newblock {\em Geometriae Dedicata}, 105(1):107--120, 2004.

\bibitem{flory_molecular_1941}
Paul~J Flory.
\newblock Molecular size distribution in three dimensional polymers. {I}.
  {Gelation}1.
\newblock {\em Journal of the American Chemical Society}, 63(11):3083--3090,
  1941.

\bibitem{garboczi_geometrical_1995}
E.~J. Garboczi, K.~A. Snyder, J.~F. Douglas, and M.~F. Thorpe.
\newblock Geometrical percolation threshold of overlapping ellipsoids.
\newblock {\em Physical Review E}, 52(1):819--828, July 1995.

\bibitem{grassberger_critical_2003}
Peter Grassberger.
\newblock Critical percolation in high dimensions.
\newblock {\em Physical Review E}, 67(3), March 2003.

\bibitem{hadwiger_vorlesungen_1957}
H.~Hadwiger.
\newblock {\em Vorlesungen über {Inhalt}, {Oberfläche} and {Isoperimetrie}}.
\newblock Springer Berlin, 1957.

\bibitem{hazanov_order_1994}
S.~Hazanov and C.~Huet.
\newblock Order relationships for boundary conditions effect in heterogeneous
  bodies smaller than the representative volume.
\newblock {\em Journal of the Mechanics and Physics of Solids},
  43(42):1995--2011, 1994.

\bibitem{huet_application_1990}
C.~Huet.
\newblock Application of variational concepts to size effects in elastic
  heterogeneous bodies.
\newblock {\em Journal of the Mechanics and Physics of Solids},
  21(43):813--841, 1990.

\bibitem{hunt_upscaling_1998}
A.G. Hunt.
\newblock Upscaling in {Subsurface} {Transport} {Using} {Cluster} {Statistics}
  of {Percolation}.
\newblock {\em Transport in Porous Media}, 30(2):177--198, 1998.

\bibitem{kesten_percolation_1982}
Harry Kesten.
\newblock {\em Percolation {Theory} for {Mathematicians}}.
\newblock Birkhäuser Boston, Boston, MA, 1982.

\bibitem{matheron_random_1975}
G.~Matheron.
\newblock {\em Random sets and integral geometry}.
\newblock Wiley series in probability and mathematical statistics:
  {Probability} and mathematical statistics. Wiley, 1975.

\bibitem{mecke_morphology_1997}
K.~R. Mecke.
\newblock Morphology of spatial patterns --- porous media, spinodal
  decomposition and dissipative structures.
\newblock {\em Acta Physica Polonica B}, 28(8):1747 -- 1782, 1997.

\bibitem{mecke_euler_1991}
K.R. Mecke and H.~Wagner.
\newblock Euler characteristic and related measures for random geometric sets.
\newblock {\em Journal of Statistical Physics}, 64(3-4):843--850, 1991.

\bibitem{nachtrab_beyond_2015}
Susan Nachtrab, Matthias J~F Hoffmann, Sebastian~C Kapfer, Gerd~E
  Schröder-Turk, and Klaus Mecke.
\newblock Beyond the percolation universality class: the vertex split model for
  tetravalent lattices.
\newblock {\em New Journal of Physics}, 17(4):043061, April 2015.

\bibitem{neher_topological_2008}
Richard~A. Neher, Klaus Mecke, and Herbert Wagner.
\newblock Topological estimation of percolation thresholds.
\newblock {\em Journal of Statistical Mechanics: Theory and Experiment},
  2008(01):P01011, 2008.

\bibitem{neuweiler_upscaling_2007}
Insa Neuweiler and Hans-Jörg Vogel.
\newblock Upscaling for unsaturated flow for non-{Gaussian} heterogeneous
  porous media.
\newblock {\em Water Resources Research}, 43(3), March 2007.

\bibitem{okun_euler_1990}
B.~L. Okun.
\newblock Euler {Characteristic} in {Percolation} {Theory}.
\newblock {\em Journal of Statistical Physics}, 59(1/2):523--527, 1990.

\bibitem{rintoul_precise_1997}
M~D Rintoul and S~Torquato.
\newblock Precise determination of the critical threshold and exponents in a
  three-dimensional continuum percolation model.
\newblock {\em Journal of Physics A: Mathematical and General},
  30(16):L585--L592, August 1997.

\bibitem{roubin_meso-scale_2013}
Emmanuel Roubin.
\newblock {\em Meso-scale {FE} and morphological modeling of heterogeneous
  media: application to cementitious materials}.
\newblock PhD thesis, École normal Supérieure de Cachan, LMT-Cachan, 2013.

\bibitem{roubin_meso-scale_2015}
Emmanuel Roubin, Jean-Baptiste Colliat, and Nathan Benkemoun.
\newblock Meso-scale modeling of concrete: {A} morphological description based
  on excursion sets of {Random} {Fields}.
\newblock {\em Computational Materials Science}, 102:183--195, May 2015.

\bibitem{scher_critical_1970}
H.~Scher and R.~Zallen.
\newblock Critical density in percolation process.
\newblock {\em J. Chem. Phys.}, 53:3759, 1970.

\bibitem{schneider_stochastic_2008}
Rolf Schneider and Wolfgang Weil.
\newblock {\em Stochastic and integral geometry}.
\newblock Probability and its applications. Springer, Berlin, 2008.

\bibitem{skal_percolation_1973}
A.~S. Skal, B.~I. Shkloviskii, and A.~L. Efros.
\newblock Percolation level in a three-dimensional random potential.
\newblock {\em JETP Lett.}, 17:377, 1973.

\bibitem{stauffer_introduction_1994}
Dietrich Stauffer and Ammon Aharony.
\newblock {\em Introduction to percolation theory}.
\newblock CRC press, 1994.

\bibitem{sykes_exact_1964}
M.~F. Sykes and J.~W. Essam.
\newblock Exact {Critical} {Percolation} {Probabilities} for {Site} and {Bond}
  {Problems} in {Two} {Dimensions}.
\newblock {\em Journal of Mathematical Physics}, 5(8):1117, 1964.

\bibitem{taylor_gaussian_2006}
Jonathan~E. Taylor.
\newblock A {Gaussian} kinematic formula.
\newblock {\em The Annals of Probability}, 34(1):122--158, February 2006.

\bibitem{taylor_euler_2003}
Jonathan~E. Taylor and Robert~J. Adler.
\newblock Euler characteristics for {Gaussian} fields on manifolds.
\newblock {\em Ann. Probab.}, 31(2):533--563, 2003.

\bibitem{tomita_percolation_1994}
H~Tomita and C~Murakami.
\newblock Percolation {Pattern} in {Continuous} {Media} and {Its} {Topology}.
\newblock In {\em Research of {Pattern} {Formation}}, pages 197 -- 203. R.
  Takaki, ktk scientific publishers edition, 1994.

\bibitem{torquato_effect_2012}
S.~Torquato and Y.~Jiao.
\newblock Effect of dimensionality on the continuum percolation of overlapping
  hyperspheres and hypercubes. {II}. {Simulation} results and analyses.
\newblock {\em The Journal of Chemical Physics}, 137(7):074106, 2012.

\bibitem{worsley_geometry_1996}
Keith~J. Worsley.
\newblock The {Geometry} of {Random} {Images}.
\newblock {\em CHANCE}, 9(1):27--40, 1996.

\bibitem{yi_analytical_2004}
Y.-B. Yi and A.~M. Sastry.
\newblock Analytical approximation of the percolation threshold for overlapping
  ellipsoids of revolution.
\newblock {\em Proceedings of the Royal Society A: Mathematical, Physical and
  Engineering Sciences}, 460(2048):2353--2380, August 2004.

\bibitem{zallen_physics_1983}
Richard Zallen.
\newblock {\em The {Physics} of {Amorphous} {Solids}}.
\newblock Wiley-VCH Verlag GmbH \& Co. KGaA, Weinheim, FRG, November 1983.

\bibitem{thorpe_sixteen-percent_2001}
Richard Zallen.
\newblock The {Sixteen}-{Percent} {Solution}: {Critical} {Volume} {Fraction}
  for {Percolation}.
\newblock In M.F. Thorpe and J.C. Phillips, editors, {\em Phase {Transitions}
  and {Self}-{Organization} in {Electronic} and {Molecular} {Networks}},
  Fundamental {Materials} {Research}, pages 37--41. Springer US, 2001.

\bibitem{zhang_microstructure-based_2011}
Mingzhong Zhang, Guang Ye, and Klaas van Breugel.
\newblock Microstructure-based modeling of water diffusivity in cement paste.
\newblock {\em Construction and Building Materials}, 25(42):2046--2052, 2011.

\end{thebibliography}

\end{document}